\documentclass[]{aa}
\usepackage{natbib}         
\usepackage{graphicx}
\usepackage{longtable}
\usepackage[switch]{lineno} 
\linenumbers\modulolinenumbers[5]
\bibpunct{(}{)}{;}{a}{}{,} 

\def\ndonnees{350}

\newcommand{\ind}[1]{_{\mathrm{#1}}}
\newcommand{\diff}{\mathrm{d}}
\newcommand{\dlog}{\mathrm{dlog}\,}

\def\Kepler{\emph{Kepler}}

\newcommand{\refeq}[1]{(\ref{#1})}

\def\numax{\nu\ind{max}}\def\nmax{n\ind{max}}
\newcommand{\np}{n}
\def\domaine{\mathcal{E}}
\def\Dnu{\Delta\nu}

\def\Dnuobs{\Delta\nu\ind{obs}}
\def\Dnurho{\nu_0}
\def\dzerol{d_{0\ell}}
\def\epsobs{\varepsilon\ind{obs}}

\def\alfa{\alpha\ind{obs}}

\def\Dnuas{\Delta\nu\ind{as}}

\def\Hmax{H\ind{max}}
\def\Bmax{B\ind{max}}
\def\Amax{A\ind{max}}
\def\visib{V^2}
\def\visiun{\visib_1}
\def\viside{\visib_2}

\def\lmax{{\ell\ind{max}}}

 \def\dR{\delta R}

\def\Rrho{R_0}

\newcommand\Teff{{T\ind{eff}}}

\newcommand\Ks{K\ind{S}}\newcommand\Kn{K_n}

\newcommand\coefDnuP{p_n}
\newcommand\xKT{\langle k\rangle} 
\newcommand\coefT{t} 
\newcommand\coefDnu{\gamma}

\newcommand\SO{S07}

\begin{document}
\title{Period-luminosity relations in evolved red giants explained by solar-like oscillations}
\titlerunning{Period-luminosity relations in red giants}
\author{B. Mosser\inst{1}\and
W.A. Dziembowski\inst{2,3}\and
K. Belkacem\inst{1}\and
M.J.Goupil\inst{1}\and
E. Michel\inst{1}\and
R. Samadi\inst{1}\and
I. Soszy\'nski\inst{2}\and
M. Vrard\inst{1}\and
Y. Elsworth\inst{4}\and
S. Hekker\inst{5}\and
S. Mathur\inst{6}
} \offprints{B. Mosser}

\institute{LESIA, Observatoire de Paris, CNRS, UPMC, Universit\'e
Paris-Diderot, 92195 Meudon cedex, France;
\email{benoit.mosser@obspm.fr}
\and Warsaw University Observatory, Aleje Ujazdowskie 4, 00-478 Warsaw, Poland
\and Copernicus Astronomical Center, ul. Bartycka 18, 00-787 Warsaw, Poland
\and School of Physics and Astronomy, University of Birmingham, Edgbaston, Birmingham B15 2TT, United Kingdom
\and Max Planck Institut f\"ur Sonnensystemforschung, Max Planck Strasse 2, Katlenburg-Lindau, Germany.
\and Space Science Institute, 4750 Walnut street Suite\# 205, Boulder, CO 80301, USA.
}

\abstract{Solar-like oscillations in red giants have been
investigated with the space-borne missions CoRoT and \Kepler,
while pulsations in more evolved M giants have been studied with
ground-based microlensing surveys. After 3.1 years of observation
with \Kepler, it is now possible to link these
different observations of semi-regular variables.}
%
{We aim to identify period-luminosity sequences in evolved red
giants identified as semi-regular variables and to interpret them
in terms of solar-like oscillations. Then, we investigate the
consequences of the comparison of ground-based and space-borne
observations.}
{We first measured global oscillation parameters of evolved
red giants observed with \Kepler\ with the envelope
autocorrelation function method. We then used an extended form of
the universal red giant oscillation pattern, extrapolated to very
low frequency, to fully identify their oscillations. The
comparison with ground-based results was then used to express the
period-luminosity relation as a relation between the large
frequency separation and the stellar luminosity.}
{From the link between red giant oscillations observed by \Kepler\
and period-luminosity sequences, we have identified these
relations in evolved red giants as radial and non-radial
solar-like oscillations. We were able to expand scaling relations
at very low frequency (periods as long as 100 days and large
frequency separation less than 0.05\,$\mu$Hz). This helped us
identify the different sequences of period-luminosity relations,
and allowed us to propose a calibration of the K magnitude with
the observed large frequency separation.}
{Interpreting period-luminosity relations in red giants in terms
of solar-like oscillations allows us to investigate the time
series obtained from ground-based microlensing surveys with a firm
physical basis. This can be done with an analytical expression
that describes the low-frequency oscillation spectra. The
different behavior of oscillations at low frequency, with
frequency separations scaling only approximately with the square
root of the mean stellar density, can be used to precisely address
the physics of the semi-regular variables. This will allow
improved distance measurements and opens the way to extragalactic
asteroseismology with the observations of M giants in the
Magellanic Clouds.}


\maketitle

\voffset = 1.2cm

\section{Introduction\label{introduction}}


The space-borne missions CoRoT and \Kepler\ have provided many
observations and carried out important results, depicting
oscillations in red giants as solar-like
\citep{2009Natur.459..398D,2010ApJ...713L.176B}. The oscillation
spectra are understood well, including the coupling of waves
sounding the core
\citep{2011Sci...332..205B,2011Natur.471..608B,2011A&A...532A..86M,2012A&A...540A.143M}
or the effects of rotation
\citep{2012Natur.481...55B,2012ApJ...756...19D,2012A&A...548A..10M,2013A&A...549A..75G,2013A&A...549A..74M}.
These results mostly concern red giants on the low red giant
branch (RGB) and on the red clump.

From the ground, microlensing surveys such as MACHO or OGLE have
provided a wealth of information on the pulsations observed in M
giants \citep[e.g.,][and references
therein]{1999IAUS..191..151W,2004MNRAS.349.1059W,2007AcA....57..201S,2010MNRAS.409..777T,2013ApJ...763..103S}.
These M giants have larger radii than the ones observed with
\Kepler. The nature of their pulsations has been questioned for
some time. There is a fundamental question whether they are
self-excited pulsations or stochastically excited modes
\citep{2001MNRAS.328..601D,2001ApJ...562L.141C}. Evidence of the
link between the observed pulsations in M giants and the
solar-like oscillations detected in the K giants has been found
\citep[][hereafter DS10]{2010A&A...524A..88D}.
\cite{2010MNRAS.409..777T} state that the M giants with the
shortest periods bridge the gap between G and K giant solar-like
oscillations and M-giant pulsations, revealing a smooth continuity
on the giant sequence. We intend to reexamine this
question in detail with \Kepler\ data, for instance, through examining
the scaling relations governing global seismic parameters.

For red giants, most of the results of the microlensing surveys
are expressed as period-luminosity (PL) relations found for
different sequences \citep[e.g.,][hereafter
\SO]{2007AcA....57..201S}. Recently, \cite{2013MNRAS.431.3189T}
have compared observed period ratios to modeled values; they found
close agreement that allows them to identify PL sequences. With
long time series recorded by \Kepler, we can now also investigate
the PL relations of pulsating M giants using these data and
compare them with solar-like oscillations. Another puzzling
question concerns the degree of the observed pulsation: radial,
non-radial, or both? Again, \Kepler\ data may provide useful
insights and methods developed for weighting the relative
contributions of the different angular degrees and measuring the
mode visibility \citep{2012A&A...537A..30M} can be extrapolated
for M giants.

In this work, we aim to analyze oscillations at very low
frequency, with large frequency separations as low as
0.1\,$\mu$Hz, corresponding to pulsations with periods up to 100
days. We intend to extrapolate the results previously obtained for
less-evolved RGB stars to the low-frequency domain where M giants
oscillate. In Section \ref{data}, we present \Kepler\ data that
are used and the tools for analyzing them. A similar presentation
of the current status of OGLE small amplitude red giants (OSARG)
is done in Section \ref{dataOGLE}. In Section \ref{scaling},
different scaling relations of global oscillation parameters are
used to verify that oscillations at very low frequency behave like
solar-like oscillations. For fully identifying the oscillations,
we need to extrapolate the findings of \cite{2011A&A...525L...9M},
who have proposed a method providing an analytical description of
the red giant oscillation pattern, based on homology
consideration. The method presented in Section \ref{individual}
allows us, for the first time, to identify the radial order and
the angular degree of solar-like oscillations in M giants. In
Section \ref{PL-identification}, we show how the low-frequency
oscillation pattern coincides with PL sequences.
This allows a precise physical interpretation, as well as a
precise calibration of the PL relation (Section
\ref{PL-calibration}). In Section \ref{discussion}, we reanalyze
previous results with the new findings. We also investigate the
possibility of enhancing the accuracy of distance measurements
using solar-like oscillations in giants.

\begin{figure}
\includegraphics[width=8.8cm]{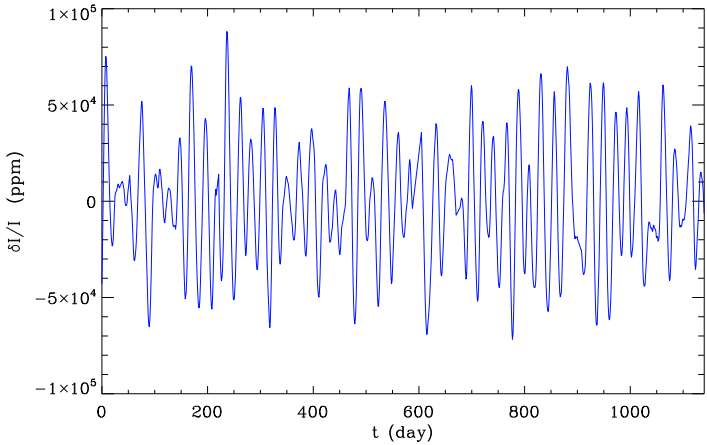}
\caption{Typical time series of the relative photometric variation
of an evolved red giant observed by \Kepler\ (KIC 2831290). The
relative  amplitude is expressed in ppm. \label{timeseries}}
\end{figure}

\section{\Kepler\ data and analysis\label{data}}

\subsection{\Kepler\ data}

We used \Kepler\ long-cadence data recorded up to and including
the \Kepler\ observing run Q13, which correspond to the targets
considered by the \Kepler\ red giant working group \citep[see,
e.g.,][]{2010ApJ...713L.176B}. The 38-month-long observation time
span provides a frequency resolution of about 10\,nHz. Compared to
previous work on \Kepler\ red giants, the time series benefited
from a refined treatment. The light curves have been extracted
from the pixel data, following the methods and corrections
described in \cite{2011MNRAS.414L...6G} and in Bloemen et al.
(2013, in preparation). In a large number of times series, this
dedicated treatment provides evidence of large irregular
low-frequency variations. It then allows the investigation of
oscillations of stars on the upper RGB and on the AGB. Contrary to
microlensing studies that deliver a limited number of periods,
often only one \citep[e.g.,][]{2005AJ....129..768F}, up to three
\citep[e.g.,][hereafter T13]{2013MNRAS.431.3189T}, or up to four
\citep[e.g.,][]{2010MNRAS.409..777T}, we aim to analyze all peaks
that can be identified as reliable in a Fourier spectrum which is
free of any aliasing effect.

\subsection{Analysis with the envelope autocorrelation function}

We used the envelope autocorrelation function
\citep{2009A&A...508..877M} to measure the observed large
separation. The method is efficient at low frequency: the envelope
autocorrelation function (EACF) corresponds to the autocorrelation
of the time series (Fig.~\ref{timeseries}), with a clear signal at
very low frequency, despite a low response due to the small number
of frequency bins in the frequency range where the oscillation
signal is seen. The precision is, however, limited by the small
number of observable modes and the relatively poor frequency
resolution. Simulations indicate a relative uncertainty of 5\,\%
for large frequency separation of 0.1\,$\mu$Hz measured with the
EACF method (compared to less than 1\,\% at the clump where
$\Dnuobs\simeq 4\,\mu$Hz).

Because it benefits from the long time series, almost all 1444
stars of our red giant list exhibit solar-like oscillations. The
few with no clear signal have too long periods, or too dim
magnitudes, or have complex spectra due to binarity
\citep[e.g.,][]{2013ApJ...767...82G}. We chose to restrict the
study to stars with large frequency separations below
2.5\,$\mu$Hz, corresponding to a frequency $\numax$ of maximum
oscillation signal below 20\,$\mu$Hz (oscillation periods longer
than 14~hours). With this threshold, the date set was reduced to
\ndonnees\ red giants. We therefore excluded clump stars, but kept
RGB stars with high enough large separations to have an
oscillation spectrum that is described precisely and understood
with previous work on red giant oscillations
\citep[e.g.,][]{2009Natur.459..398D,2010ApJ...723.1607H,2010A&A...517A..22M}.
The extrapolation of these properties was used to guide the
analysis of more evolved stars. Thus, large frequency separations
down to 0.047\,$\mu$Hz could be measured, even if they were not
directly accessible with the standard methods. This corresponds to
stars with a maximum oscillation signal peaking around
0.12\,$\mu$Hz (periods of about 100\,days).

\subsection{Non-asymptotic conditions}

The large separation we measure in observed spectra is different
from the theoretical asymptotic large separation. Asymptotic
conditions require observations at very high radial orders. This
condition cannot be met with high-luminosity red giants.
Therefore, following \cite{2013A&A...550A.126M} and
\cite{belkacem}, we distinguish the observed large frequency
separation $\Dnuobs$, its asymptotic counterpart $\Dnuas$, and the
dynamical frequency $\Dnurho$ scaling as $\sqrt{M/R^3}$. A priori,
$\Dnurho$ and not $\Dnuobs$ must be used in the scaling relations
that provide estimates of the stellar mass and radius. Modeling
shows that $\Dnuobs$ provides an appropriate proxy of $\Dnurho$
\citep{belkacem}.

\begin{figure}
\includegraphics[width=8.8cm]{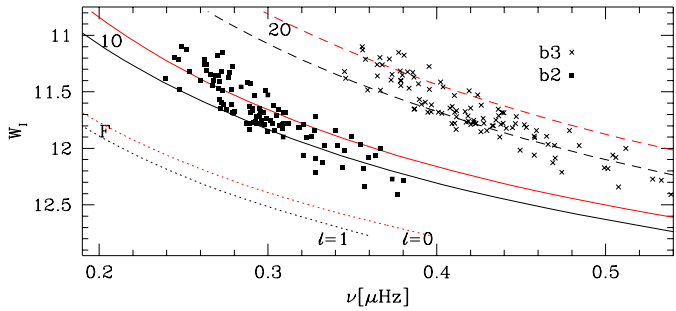}
\caption{The frequency--luminosity relations for OSARG with
dominant peaks identified as $b_2$ and $b_3$ ridges compared with
the RGB model values for the first three radial (red lines) and
dipole modes (black lines). F stands for fundamental radial
oscillation, 1O and 2O for first and second radial overtones,
respectively. \label{figOGLE}}
\end{figure}

\section{OGLE data\label{dataOGLE}}

For comparison with \Kepler\ data, we considered data from the
OGLE-III Catalog of Variable Stars. This catalog is based on
infrared photometric data collected in the Large Magellanic Cloud
(LMC) during eight years of continuous observations
\citep{2009AcA....59..239S}. It contains about $8 \times 10^4$
OSARG.

\subsection{OSARG}

The acronym OSARG has been introduced by
\cite{2004MNRAS.349.1059W} for the low-amplitude red variables,
which were detected in the OGLE microlensing survey of the
Galactic bulge. \cite{2004AcA....54..129S} identified seven
distinct sequences in the PL plane for the OSARG in both
Magellanic Clouds. The ones denoted in the decreasing period order
as a$_1$-a$_4$ were attributed to AGB and those denoted as
$b_1$-$b_3$ to RGB stars. The argument that these objects are
solar-like pulsators was put forward in S07, where the BaSTI
isochrones \citep{2006ApJ...642..797P} were used to convert
measured $W_I$ reddening-free Wesenheit functions to stellar
parameters along the upper RGB in the LMC. It was shown that close
to the tip of the RGB, the extrapolated $\numax$ falls between the
$b_2$ and $b_3$ ridges, which are the strongest. It was also
suggested that the first two radial overtones may be responsible
for the two ridges.

In a follow-up work, DS10 focused on objects that have high
signal-to-noise-ratio frequency peaks simultaneously on the
sequences $a_2$ and $a_3$, or $b_2$ and $b_3$. The latter two
sequences are compared with models in the frequency--$W_I$ diagram
(Fig.~\ref{figOGLE}). The periods were calculated for envelope
models along a 4-Gy isochrone at $Z=0.008$. This metallicity is
the standard value for the young population in the LMC. At the
distance modulus of 18.5 mag, the $W_I\in[12.5,11.0]$ range
corresponds to $M/M_\odot\in[1.23,1.20]$,
$\log(L/L_\odot)\in[3.07,3.40]$, and $\log T_{\rm
eff}\in[3.58,3.55]$. Different choices of the age and $Z$ were
considered in DS10. At $Z=0.004$, the $W_I(\nu)$ curves are
shifted downward by about 0.2 mag. Higher age, implying lower
mass, also causes a downward shift. A meaningful comparison for
$a_2$ and $a_3$ sequences with calculate values cannot be done due
to the lack of adequate models of AGB stars.

Dipole modes are also shown in Fig.~\ref{figOGLE}. These dipole
modes are perfectly trapped in the convective envelope. The
gravity  wave excited at the top radiative core is effectively
damped on its way toward the center. A small uniform shift of the
core keeps the center of stellar mass at rest. This is why the
p$_0$ dipolar mode exists. The energy loss by the gravity wave
emitted at the bottom of the envelope is very small, and thus the
chances of detecting such modes are essentially the same as the
radial modes \citep{2012A&A...539A..83D}. The plots in
Fig.~\ref{figOGLE} suggest that the $b_2$ ridge is composed of the
first radial overtone (and possibly quadrupole modes that have
intermediate frequencies). The $b_1$ and $b_3$ ridges are then
composed of modes by, respectively, one radial order lower or
higher than $b_2$.

This picture seems appealing but there is a difficulty, which is
stressed in DS10. In the whole data range, the frequency
difference between the modes in the $b_3$ and $b_2$ sequences is
smaller by some 20\,\% than the predicted frequency difference
between the radial second and first overtones, and greater by a
larger amount if, instead of radial modes, their dipolar
counterparts are considered. T13 have recently proposed to solve
the problem by interpreting the $b_2$ and $b_3$ ridges in terms of
modes one radial order higher than adopted in DS10.

The interpretation proposed by T13 has been contemplated by DS10
but has been found to be inconsistent with the $b_2$ and $b_3$
ridges in the $\log P-W_I$ plane. This interpretation also leads
to disagreement between observations and models in the Petersen
diagram if models based on stellar evolution calculations for
$1.1M_\odot$ star are used (see Figs. 5 and 10 in T13). At the
center of the $b_2$ ridge, the period in days is $\log P\simeq
1.6$ and $P_{b_3}/P_{b_2}\simeq 0.7$. The corresponding model
ratios are $P_{2\rm O}/P_{1\rm O}\simeq 0.67$ and $P_{3\rm
O}/P_{2\rm O}\simeq 0.75$, with $P_{k\rm O}$ the period of the
k-th overtone. The two values are in a good agreement with DS07,
who considered the range of masses encompassing $M=1.1\,M_\odot$.
T13 were apparently able to reproduce the observed ratios by
considering higher values of $M$ and $L$ indirectly inferred from
$W_I$. At this stage, it seems fair to consider mode
identification in OSARG as an open issue.

\subsection{OSARG seen as solar-like oscillations\label{ogle-solar-like}}

For the comparison with \Kepler\ data, a subsample of 723 stars
with two PL relations and relatively strong amplitudes were
selected among OSARG treated in DS10. For these data, the global
seismic parameters $\Dnuobs$ and $\numax$ were first crudely
estimated as follows.  We considered the frequency difference of
the two observed frequencies as a proxy of the frequency large
separation, and the weighted mean as a proxy of $\numax$. The
relative weights were given by the amplitudes of the peaks. The
calculation of the large separation assumes that only modes with
the same degree, presumably radial modes, were observed. In some
cases, outliers were seen with a measured frequency interval
corresponding to two times the large separation. We then used half
the measured value.

\begin{figure*}
\centering
\includegraphics[width=14.8cm]{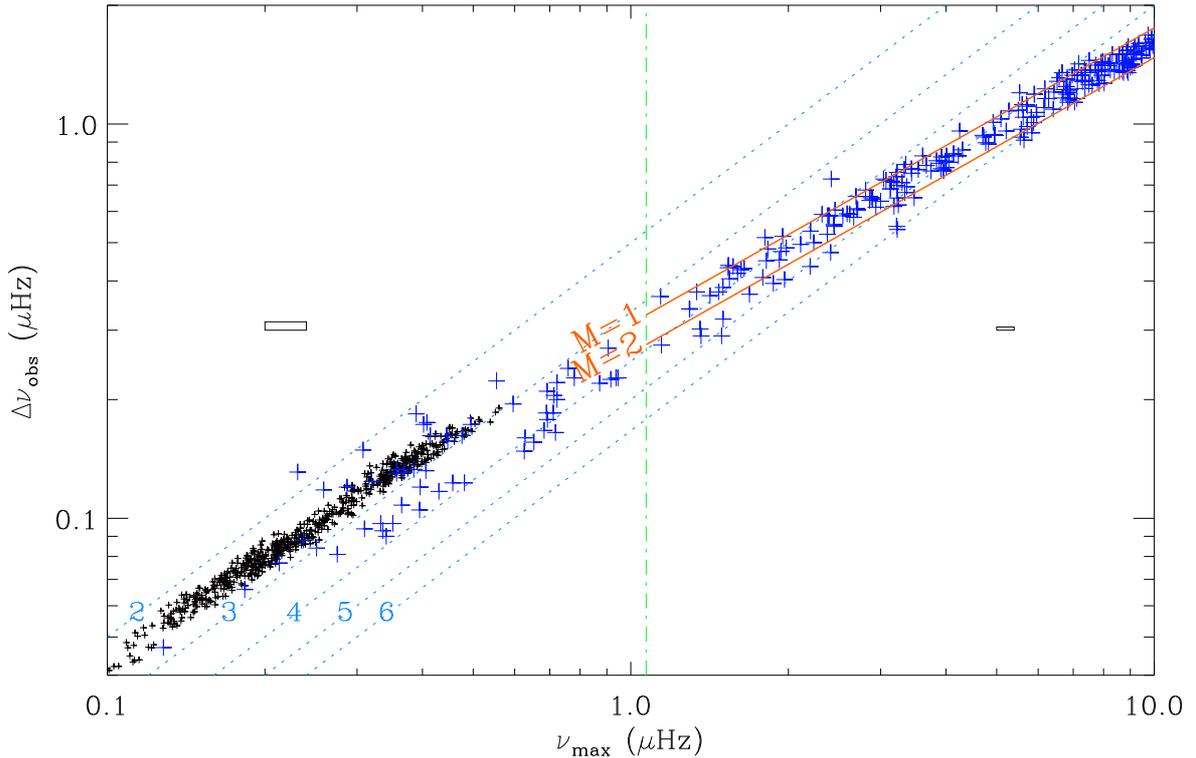}
\caption{$\numax$ -- $\Dnuobs$ relation for \Kepler\ (blue pluses)
and OGLE (black symbols) AGB and RGB stars. The red solid lines
indicate the isomass levels 1 and 2\,$M_\odot$, in their domain of
validity. The dotted lines indicate isolines of the number $\nmax
= \numax / \Dnuobs$. The vertical green dot-dashed line shows the
region where the oscillation regime changes. Typical uncertainties
at low and large $\numax$ for \Kepler\ data are indicated by the
1-$\sigma$ error boxes. \label{dnu-numax}}
\end{figure*}

\section{Global oscillation parameters\label{scaling}}

\subsection{Measuring $\numax$ at low frequency}

For the location of the excess power in \Kepler\ data, we first
assume that the excitation of the modes is stochastic. It thus
makes sense to measure the frequency $\numax$ of the maximum
oscillation signal. This hypothesis will have to be discussed a
posteriori. Measuring $\numax$ accurately at low frequency is
challenging, because $\numax$ cannot be well defined when only two
or three modes are observed. Moreover, as discussed by
\cite{2012sf2a.conf..173B}, we lack a precise theoretical
definition of $\numax$. Methods usually used are operative in this
low-frequency domain, but with large uncertainties
\citep[e.g.][]{2009A&A...508..877M,2010MNRAS.402.2049H,2010A&A...522A...1K,2011ApJ...743..143H}.
We used a very simple approach that does not rely on a fit of the
oscillation excess power. A first estimate of $\numax$ is derived
as a function of the large separation $\Dnuobs$ derived with the
EACF according to the scaling relations found at larger $\Dnuobs$
\citep{2011ApJ...743..143H}. It is used to define the frequency
range $\domaine$ with oscillation power in excess. We then perform
the weighted mean
\begin{equation}\label{mesure-numax}
\numax =  \int_\domaine \bigl[p(\nu)-b(\nu)\bigr]\; \nu\; \diff
\nu \Bigm / \int_\domaine \bigl[p(\nu)-b(\nu)\bigr]\; \diff \nu ,
\end{equation}
where $p(\nu)$ is the power spectrum density observed in the
frequency range $\domaine$, and $b(\nu)$ is the background
component, determined with the method COR presented in
\cite{2011ApJ...741..119M}. This method describes the background
in the vicinity of the frequency range where the oscillation
excess power is observed, at frequencies just below $0.4\,\numax$
and just above $1.6\,\numax$. Monte-Carlo simulations indicate
that the precision of the measurement of $\numax$ with
Eq.~\refeq{mesure-numax} is about 15\,\% for a large separation of
0.1\,$\mu$Hz (20\,\% in similar conditions when $\numax$ is
derived from a Gaussian fit of the smoothed spectrum). Reducing
these large uncertainties with longer time series could both lower
the influence of the stochastic excitation and enhance the
frequency resolution of oscillation spectra.

In the following, we use $\numax$ for linking with observations in
the earlier stages of the RGB. However, for precise information,
we prefer to use $\Dnuobs$. The $\numax$ -- $\Dnuobs$ relation
\cite[e.g.,][]{2009A&A...506..465H,2009MNRAS.400L..80S} is shown
in Fig.~\ref{dnu-numax}. The measurements at low $\Dnuobs$ extend
the trend at larger $\Dnuobs$, so that we have a first indication
that the hypothesis of stochastic excitation is verified. We
notice, however, a change of regime around 1\,$\mu$Hz, associated
with an apparent lack of data. The fits for \Kepler\ stars are,
below and above 1\,$\mu$Hz,
\begin{eqnarray}
      \Dnuobs &=& (0.296 \pm 0.004) \, \numax^{0.727\pm 0.007} \ \hbox{ for } \numax > 1 \,\mu\hbox{Hz}, \label{dnu-numax-high}\\
      \Dnuobs &=& (0.248 \pm 0.010) \, \numax^{0.783\pm 0.049} \ \hbox{ for } \numax < 1 \,\mu\hbox{Hz}, \label{dnu-numax-low}
\end{eqnarray}
with both $\Dnuobs$ and $\numax$ expressed in $\mu$Hz
(Fig.~\ref{dnu-numax}).

We note that, at low $\numax$, the OGLE and \Kepler\ relations
fully agree, with a slope slightly steeper than for higher
frequencies. \Kepler\ data show a wider spread, which we interpret
as due to the poorer frequency resolution and to the imprecise
measurement of $\numax$. The exponent remains close to three
quarters, as a direct consequence of the scaling relations of the
stellar mass and radius. At large $\Dnuobs$, the spread of the
value around the mean fit indicated by Eq.~\refeq{dnu-numax-high}
is mainly a mass effect. At low $\Dnuobs$, the spread is also due
to the inaccurate measurement of $\numax$. Compared to previous
studies \citep[e.g.,][]{2011ApJ...743..143H} we have gained more
than one decade towards low frequencies in determining the
validity of the $\numax$ -- $\Dnuobs$ relation. From this
relation, we can derive the radial order $\nmax$ corresponding
approximately to $\numax / \Dnuobs$. As already noted by
\cite{2010A&A...517A..22M}, this index decreases significantly
when the stellar radius increases.

The change in slope noted in the  $\numax$ -- $\Dnuobs$ relation
(Fig.~\ref{dnu-numax}) occurs in a frequency range with an
apparent lack of data. The density of stars being low, we cannot
exclude that this gap is only spurious. However, we do not
identify any observing bias able to produce such a gap. Then, in
the absence of direct explanation, we have chosen to identify it
in all figures, aiming to find an explanation.

\begin{figure}
\includegraphics[width=8.8cm]{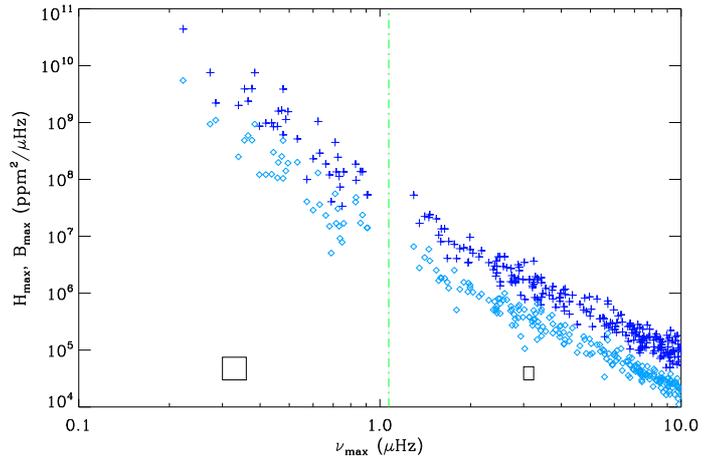}
\caption{Maximum height $\Hmax$ (blue $+$) and background power
density $\Bmax$ (diamonds) at $\numax$, as a function of $\numax$.
The vertical green dot-dashed line shows the region where the
oscillation regime changes. Typical uncertainties at low and large
$\numax$ are indicated by the 1-$\sigma$ error boxes.
\label{fig-Hmax}}
\end{figure}

\begin{figure}
\includegraphics[width=8.8cm]{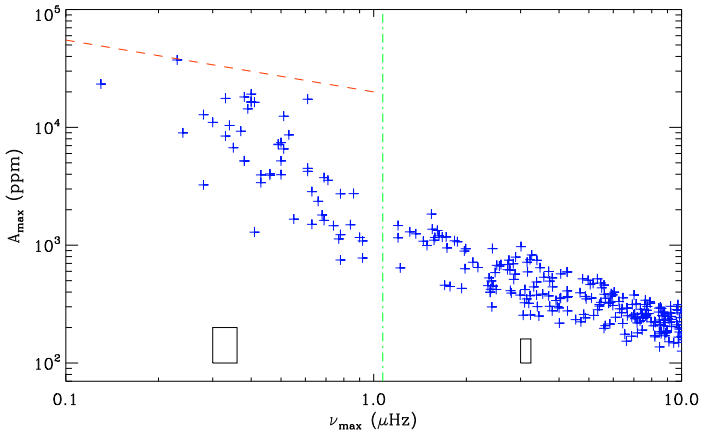}
\caption{Same as Fig. \ref{fig-Hmax} but for the maximum relative
amplitude $\Amax$. The red dashed line shows the relative
amplitude that causes an acceleration as strong as the surface
gravity, according to the model developed in Section
\ref{mass_loss}. \label{fig-Amax}}
\end{figure}

\subsection{Background, maximum height}

The maximum height of the power density spectrum at $\numax$, the
background at $\numax$, and the maximum amplitude were derived
from the method COR used in \cite{2012A&A...537A..30M} and
\cite{2011ApJ...741..119M}. The variation in these parameters as a
function of $\numax$ is presented in Fig.~\ref{fig-Hmax}. The
scaling relations, for height and background in
ppm$^2\,\mu$Hz$^{-1}$ and $\numax$ in $\mu$Hz, are
\begin{eqnarray}
  \Hmax  &=& (2.8\pm0.2)\, 10^7 \ \numax^{-2.50\pm 0.05} ,  \\
  \Bmax  &=& (6.2\pm0.4)\, 10^6 \ \numax^{-2.48\pm 0.04}  \ \hbox{ for } \numax > 1 \,\mu\hbox{Hz}
\end{eqnarray}
and
\begin{eqnarray}
  \Hmax  &=& (3.6\pm1.1)\, 10^7 \ \numax^{-4.10\pm 0.41} , \\
  \Bmax  &=& (6.3\pm1.3)\, 10^6 \ \numax^{-3.35\pm 0.30}  \ \hbox{ for } \numax < 1 \,\mu\hbox{Hz}.
\end{eqnarray}
For $\numax > 1 \,\mu$Hz, the values of the fits are comparable to
the values found by \cite{2012A&A...537A..30M} in the lower RGB
and in the clump. Interestingly, the exponent of the $\Hmax$
scaling relation is close to the exponent found in the RGB by
\cite{2012A&A...543A.120S} for the energy supply rate, varying as
$(L/M)^{2.6}$, hence as $\numax^{-2.6}$. This may have indirect
consequences on the mode linewidth.

At very low frequency, below 0.25\,$\mu$Hz, we have identified in
a few cases a component that may be either a stellar signal
(activity or background modulated by the surface rotation) or
instrumental noise. This effect occurs at such a low frequency
that it only modestly perturbs the measurement of the background
of the most evolved stars, without significant consequences.

Despite the change of regime, which occurs at the same location as
for the $\numax$ -- $\Dnuobs$ scaling relation, the continuous
variation from high to low values of $\Dnuobs$ confirms the
hypothesis of stochastic oscillations. The lack of stars with
global parameters $\numax \simeq 1\,\mu$Hz, or equivalently
$\Dnuobs \simeq 0.28\,\mu$Hz, and the change of regime in all
scaling relations tend to indicate a change in either the stellar
structure or the upper atmospheric properties. However, in both
regimes, the exponents for $\Hmax(\numax)$ and $\Bmax(\numax)$ are
consistent. As for earlier evolutionary stages, the ratio $\Hmax /
\Bmax$ does not vary with $\numax$. This underlines that the ratio
of the energy transferred from the convection in the granulation
(background signal) or in the oscillation is nearly constant.

\subsection{Maximum amplitude\label{ampmax}}

To prepare the link with infrared ground-based data, we also
investigate the mode amplitude $\Amax$ of the oscillations. This
parameter can be derived from $\Hmax$, with the method presented
in \cite{2012A&A...537A..30M}:
\begin{eqnarray}
  \Amax  &=& (1.4\pm0.1)\; 10^3 \; \numax^{-0.82\pm 0.03}  \ \hbox{ for } \numax > 1 \,\mu\hbox{Hz} , \\
  \Amax  &=& (2.0\pm0.3)\; 10^3 \; \numax^{-1.53\pm 0.21}  \ \hbox{ for } \numax < 1 \,\mu\hbox{Hz}, \label{eqt-Amax}
\end{eqnarray}
with $\Amax$ in ppm and $\numax$ in $\mu$Hz. At high $\numax$, we
retrieve a steeper fit than previously found for the lower part of
the RGB (slope of $\-0.71\pm0.01$). At low $\numax$, $\Amax$ can
also be directly measured in the time series since relative
photometric variations are greater than 1\,mmag for red giants
with $\Dnu \le 0.4\,\mu$Hz (Fig.~\ref{fig-Amax}). Accordingly, we
have estimated the mean amplitude $\sigma_t$ of the oscillating
signal: it directly compares to $\Amax$. This extends the validity
of the relation found in the RGB by \cite{2012A&A...544A..90H}.
Extrapolation of Eq.~\refeq{eqt-Amax} predicts $\Amax = 0.75 $ for
modes with $\numax\simeq 0.02\,\mu$Hz (periods about 580 days),
hence a peak-to-valley visible magnitude change of 1.

We did not consider superimposing the OGLE data on the \Kepler\
data in Figs.~\ref{fig-Hmax} and \ref{fig-Amax} since the
different ways the data have been treated preclude a direct
comparison. In fact, DS10 already perform a comparison, based on
the amplitude measure in the I band, which agrees with the scaling
relations found for RGB stars observed by CoRoT and \Kepler\
\citep{2010A&A...517A..22M,2010ApJ...713L.182S}.

\begin{figure}
\includegraphics[width=8.3cm]{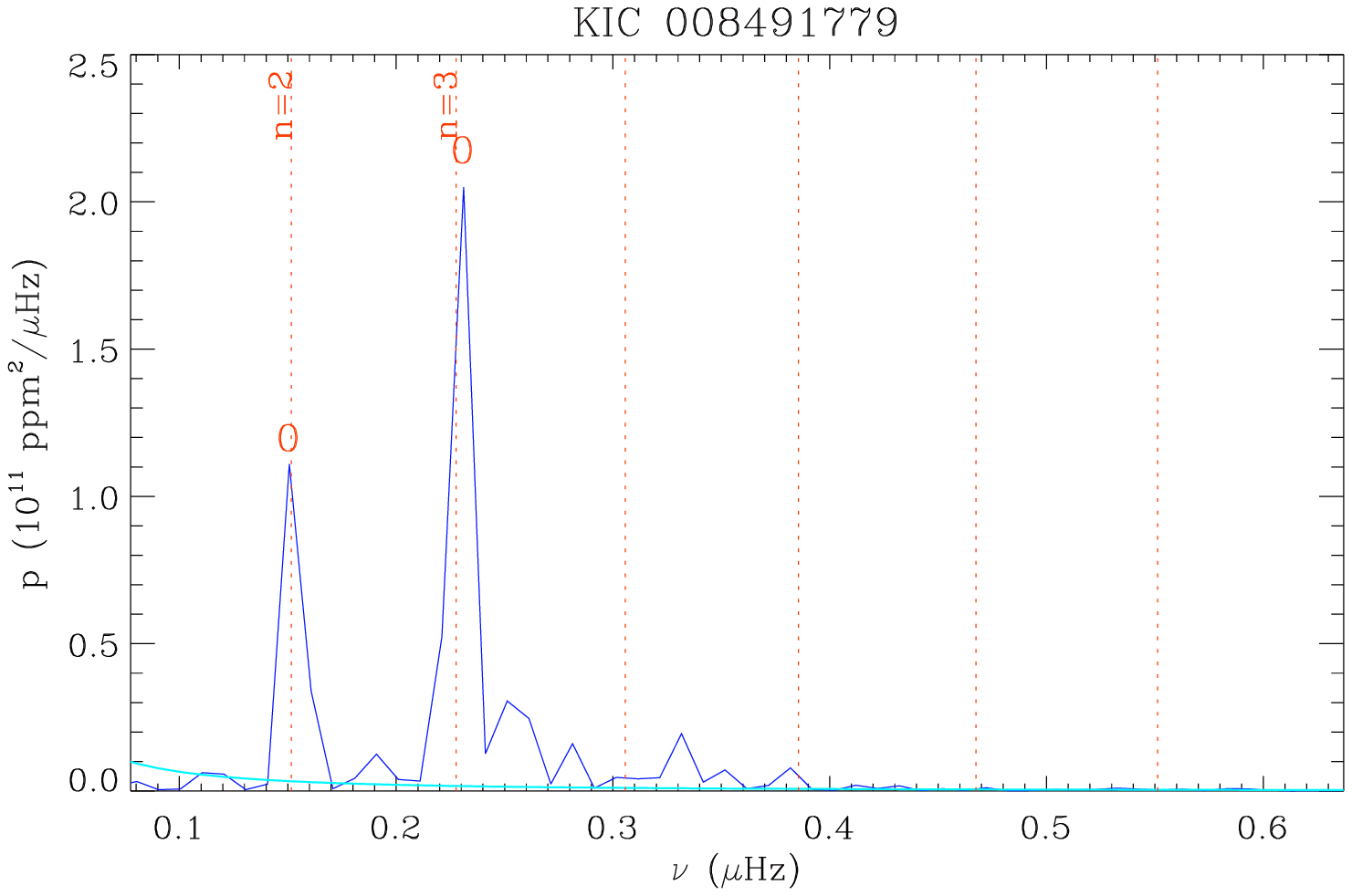}
\includegraphics[width=8.3cm]{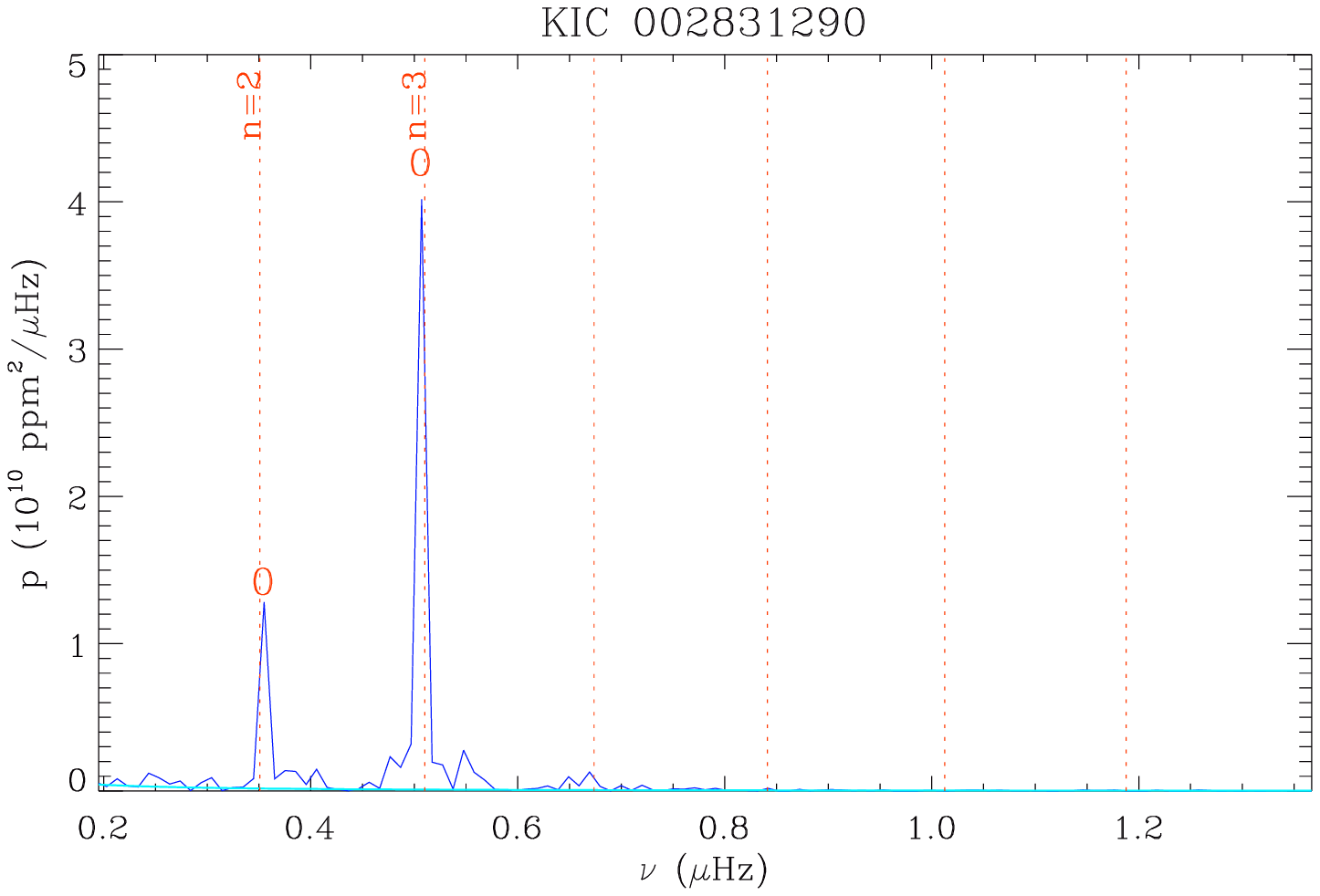}
\includegraphics[width=8.3cm]{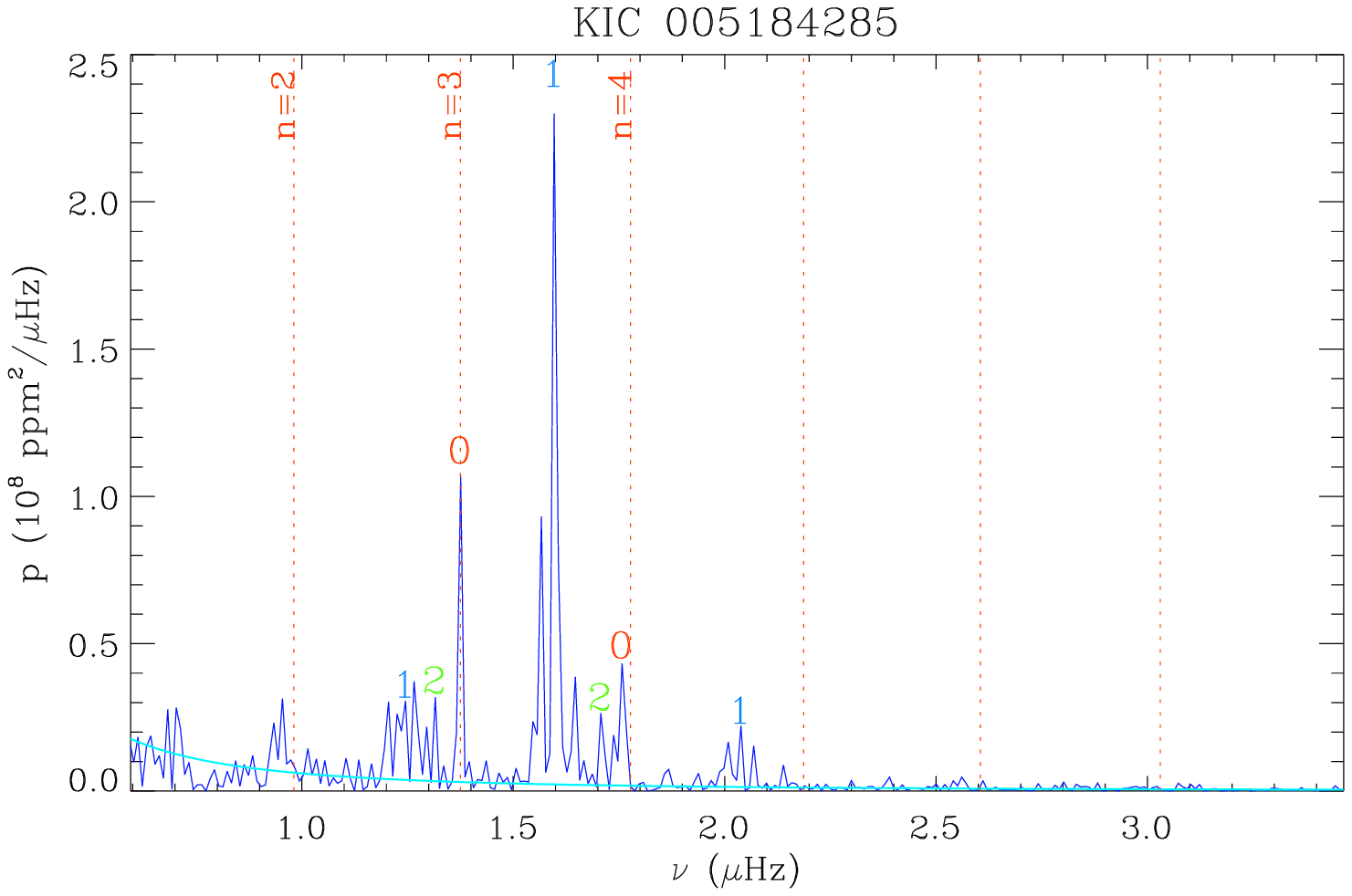}
\includegraphics[width=8.3cm]{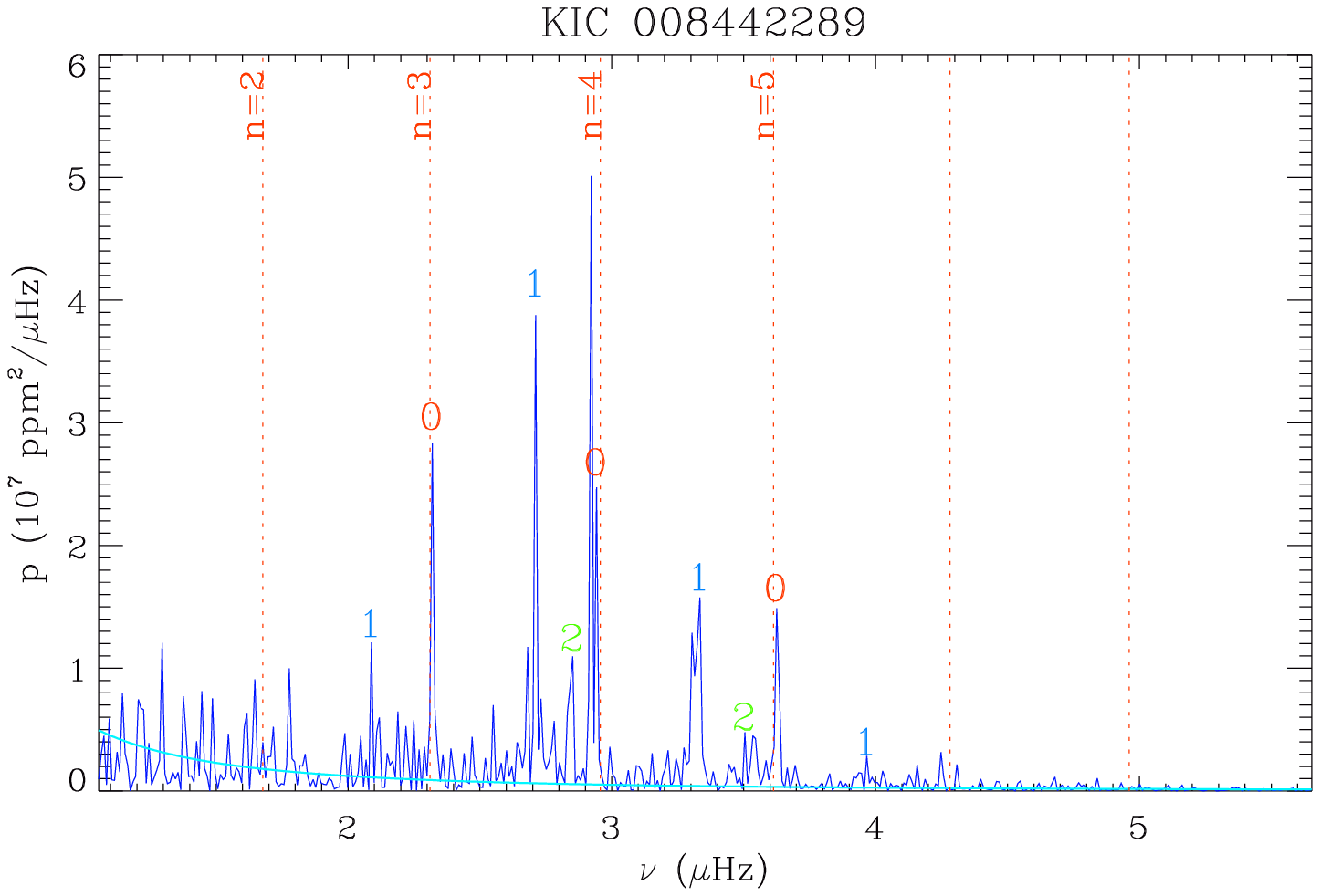}
\caption{Typical \Kepler\ spectra with low $\Dnuobs$. The mode
identification is derived from the method presented in Section
\ref{individual}. Dotted lines indicate the expected location of
radial modes. Modes at least eight times larger than the
background (light-blue line) are labeled with their degree. The
presence of multiple peaks corresponding to dipole modes indicates
that they are mixed modes. \label{fig0}}
\end{figure}

\begin{figure}
\includegraphics[width=8.8cm]{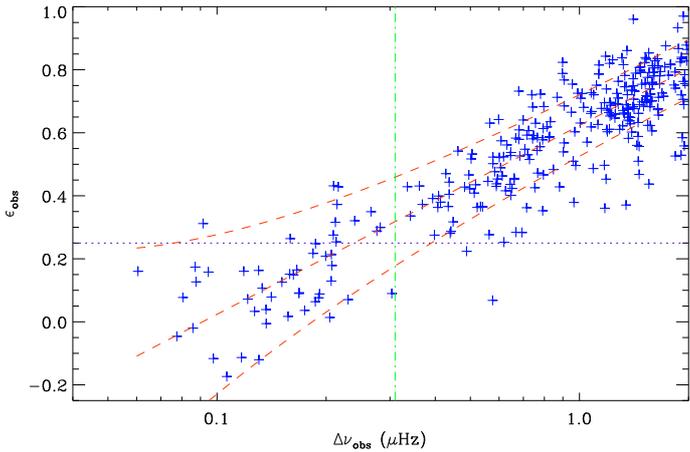}
\caption{Relation $\epsobs (\Dnuobs)$ at low frequency. The dashed
lines correspond to the fit proposed in Table \ref{tab_new_fit},
plus and minus the uncertainty, which mainly reflects the
uncertainty on $\Dnuobs$.
The horizontal dotted line shows the
asymptotic value $\varepsilon\ind{as} = 1/4$.\label{fig-dnu-eps}}
\end{figure}

\begin{figure}
\includegraphics[width=8.8cm]{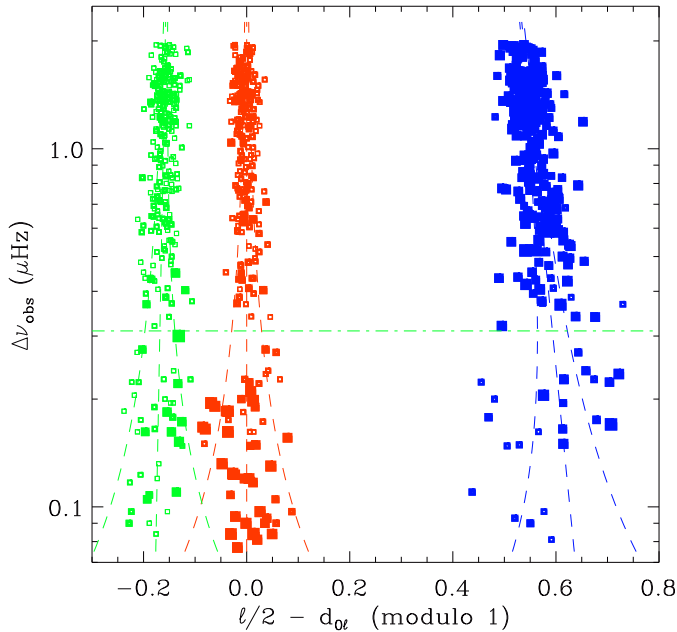}
\caption{Variation in the terms $\ell/2 - \dzerol$ measuring the
location of non-radial modes with respect to radial modes (Eq.
\ref{tassoul_obs}) in abscissa, with the observed large separation
in ordinate. The term $d_{00}$ in red, close to zero, indicates
that the fit of $\epsobs (\Dnuobs)$ is unbiased; $0.5 - d_{01}$ in
blue shows a wide spread that may indicate the presence of mixed
modes; $-d_{02}$ is plotted in green. The size of the symbols is
proportional to the percentage of energy in a given degree. The
dashed lines correspond to the fits proposed in Table
\ref{tab_new_fit}, plus or minus the frequency resolution.
\label{figuniv}}
\end{figure}

\section{Individual frequencies\label{individual}}

The mode identification for red giant oscillations observed by
CoRoT and \Kepler\ is provided in an automated way by the
universal red giant oscillation pattern
\citep{2011A&A...525L...9M}. It is based on the assumption that
the near-homology of red giant interior structure implies the
homology of the oscillation spectra, as verified by subsequent
work \citep[e.g.,][]{2012ApJ...757..190C}. The method, up to now
tested for $\Dnu \ge 0.4\,\mu$Hz and validated by comparison with
other methods
\citep{2011A&A...525A.131H,2011MNRAS.415.3539V,2012A&A...541A..51K,2012A&A...544A..90H},
has been extrapolated towards lower frequencies.

\subsection{Parametrization of the spectrum}

For red giants, it is useful to express the radial and low-degree
mode frequencies from the analytical expression
\citep{2011A&A...525L...9M}
\begin{equation}\label{tassoul_obs}
\nu_{\np,\ell} = \left(\np  + \epsobs
 + {\alfa \over 2}\; [ \np - \nmax ]^2 +{\ell \over 2} - \dzerol \right)
 \Dnuobs.
\end{equation}
The first three terms ($n$, $\epsobs$, and the second-order
correction in $\alfa$) provide the expected mean location of the
radial modes, independent of small modulations that are due to
inner-structure discontinuities
\citep[e.g.,][]{2010A&A...520L...6M}. Compared to the radial
modes, the relative positions of dipole and quadrupole modes
differ by the terms $\ell /2 -\dzerol$.

The dimensionless term $\nmax$, defined as $\numax / \Dnuobs -
\epsobs$, is introduced in Eq.~\refeq{tassoul_obs} to allow for
the fact that the large separation $\Dnuobs$ is measured in a
frequency range centered on $\numax$. The curvature term $\alfa$
expresses the second-order contribution of the asymptotic
expansion. It varies as $0.076/\nmax$ for less-evolved red giants
\citep{2013A&A...550A.126M}; this implies a significant gradient
for low $\nmax$, as high as 4\,\%, of the frequency separation
between consecutive radial modes. We note that such a gradient is
clearly observed in the spectra of M giants reported by
\cite{2010MNRAS.409..777T}, which does not necessarily correspond
to the gradient expressed by $\alfa$ since significant departure
from the asymptotic expansion is expected at very low radial
order.

We chose to express the dependence of the parameters $\epsobs$,
$\alfa$, and $\dzerol$ with a term $\log_{10} \Dnuobs$, as in
\cite{2011A&A...525L...9M}. Owing to the possibility that dipole
modes behave as mixed modes, we did not fix their position during
the analysis. We also considered that the influence of rotation is
negligible. Extrapolation from \cite{2012A&A...548A..10M}
indicates that the transfer of angular momentum between the
envelope and the core is efficient enough for ensuring very small
splittings of the dipole modes either on the upper RGB or on the
AGB.

\begin{table}
 \caption{Fits of the low-degree ridges}\label{tab_new_fit}
 \begin{tabular}{llcc}
   \hline
 & $\ell$ &\multicolumn{2}{c}{fit $A_\ell +B_\ell \log\Dnuobs$}\\
                  &  & $A_\ell$           &  $B_\ell$          \\
\hline
$\epsobs$& 0 & $ 0.623 \pm 0.006$ & $ 0.599 \pm 0.015$ \\
$d_{01}$ & 1 & $-0.057 \pm 0.003$ & $ 0.070 \pm 0.007$ \\
$d_{02}$ & 2 & $0.162 \pm 0.002$  & $-0.013 \pm 0.004$ \\
\hline
$\alfa$  &all& \multicolumn{2}{c}{$0.076 \ \Dnuobs/ \numax$}  \\
\hline
\end{tabular}

These fits are valid for observed large separations $\Dnuobs$ in
the range [0.1 -- 2\,$\mu$Hz].
\end{table}

\begin{figure}
\includegraphics[width=8.8cm]{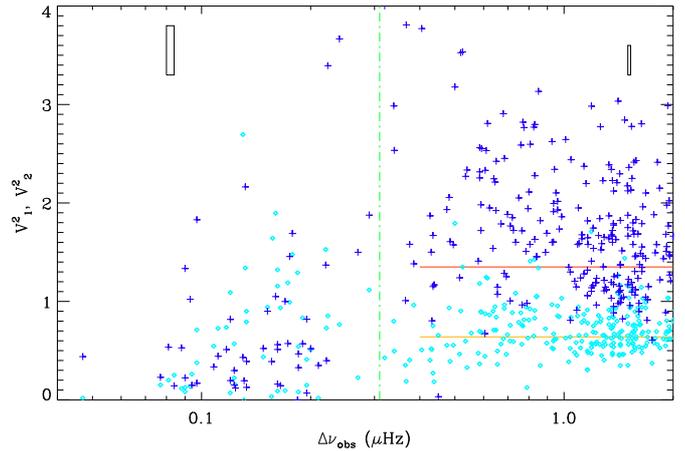}
\caption{Dipole (dark blue $+$) and quadrupole mode (light blue
diamonds) visibilities, as a function of the observed large
separation. The horizontal lines indicate the respective fits
found by \cite{2012A&A...537A..30M} at larger $\Dnuobs$. The
vertical green dot-dashed line shows the region where the
oscillation regime changes. \label{visi}}
\end{figure}

\subsection{Mode identification}

As already done by \cite{2011A&A...525L...9M} for CoRoT RGB and
red clump stars, the use of Eq.~\refeq{tassoul_obs} provides a
refined measurement of $\Dnuobs$ and allows complete
identification of all low-frequency spectra. The extrapolation
towards very low frequencies was made possible by the continuous
distribution of giants at all evolutionary stages. A few examples
are shown in Fig.~\ref{fig0}.

The efficiency of the method is proven by the capability of
fitting all major peaks of the spectra, defined by a
height-to-background ratio over eight. The parameters of the fit
were iteratively improved. This provides a small correction of the
$\epsobs ( \Dnuobs)$ relation introduced by
\cite{2011A&A...525L...9M}, as plotted in Fig.~\ref{fig-dnu-eps}.

From the radial modes actually identified with
Eq.~\refeq{tassoul_obs}, we derived new values for the large
separation and for the offset and provided a new fit of the
radial-mode oscillation pattern (Table \ref{tab_new_fit}). An
\'echelle diagram shows the efficiency of the method
(Fig.~\ref{figuniv}). All peaks with a height-to-background ratio
greater than eight are plotted. Glitches are present, with a
greater relative weight than at larger $\Dnuobs$, on the order of
$\pm0.05\Dnuobs$. There is an indication that, even at low
$\Dnuobs$, dipole modes appear to be mixed since more than one
peak per radial order is often seen. However, their period spacing
cannot be estimated since the frequency resolution is not
sufficient to resolve consecutive mixed orders. The size of the
symbols in Fig.~\ref{figuniv} is proportional to the total energy
integrated for each degree. From these measurements, we find that
dipole modes dominate at large $\Dnuobs$, whereas radial modes
dominate at low $\Dnuobs$. The change in regime occurs again in
the same region.

\cite{2013A&A...550A.126M} have shown that Eq.~\refeq{tassoul_obs}
is a form equivalent to the asymptotic expansion, based on the
observed value of the radial frequency separation, which is
different from the asymptotic value. Here, we use it at very low
radial order to provide the mode identification. An operative fit
does not imply that the oscillation pattern follows the asymptotic
expansion. In fact, a significant departure to asymptotics is seen
in the function $\epsobs$, valid for $\Dnuobs$ down to
0.1\,$\mu$Hz. According to the link between observed and
asymptotic parameters, we should have $\epsobs \ge 1/4$
\citep{2013A&A...550A.126M}. This is clearly not the case since
negative values of $\epsobs$ are found at very low $\Dnuobs$. In
fact, the deviation from the asymptotic pattern at radial orders
as low as 2 is expected.

\subsection{Mode visibilities}

The degrees of the observed modes could not, so far, be determined
from ground-based microlensing surveys. Observationally, most
often only radial modes were searched for, even if the presence of
non-radial modes was suspected. In luminous red giants, all
non-radial modes suffer high radiative losses in stellar cores.
However, for some (one for each $\ell$ per radial mode), the
efficient trapping in the envelope ensures that these losses are
negligible in the overall energy budget
\citep{2009A&A...506...57D,2012A&A...539A..83D}. The expected
oscillation spectrum then again becomes as simple as in unevolved
non-rotating stars.

The location of radial modes and the resulting full identification
of the oscillation pattern in the \Kepler\ data allows us to
measure the visibilities of the modes, depending on the degree,
with the method proposed by \cite{2012A&A...537A..30M}.
Visibilities measure the mean height of the modes associated to
each degree, calibrated to the mean height of radial modes. We did
not search for $\ell=3$ modes since insufficient frequency
resolution hampers their detection at low frequency. Our results
are given in Fig.~\ref{visi}. We identify three regimes.
\begin{itemize}
    \item For large separations over 0.6\,$\mu$Hz, the measured
visibilities $\visiun$ and $\viside$ have a mean value of about
1.5 and 0.6, respectively, in agreement with the measurements done
at high $\numax$ \citep{2012A&A...537A..30M}. Their distributions
show a wide spread, which we identify as the result of the small
numbers of significantly excited modes.
    \item  For large separations in the range [0.2
    -- 0.6\,$\mu$Hz], in the domain where we observed the change in regime of previous scaling
    relations,
    we note a huge spread of the visibility distributions.
    Again, we interpret this as a result of the stochastic excitation, amplified by the fact that only
    a limited number of modes are visible. Most of
    the oscillation energy is often concentrated in one major peak, close to $\numax$,
    which can have any degree since radial and non-radial are simultaneously visible. The degree $\lmax$ of this peak is associated to a dominant visibility $V^2_\lmax$.
    \item At lower large separations, we note a rapid decrease in the non-radial
    mode visibilities, both for dipole and quadrupole modes. For
    large separations below 0.2\,$\mu$Hz, the damping is severe, so that only radial modes subsist with
    non-negligible amplitudes.
\end{itemize}

\subsection{Large frequency separations in OGLE data\label{newOGLE_Dnu}}

Large separations of oscillation spectra observed by the OGLE
survey were obtained according to the parametrization expressed by
Eq.~\refeq{tassoul_obs}. For $\Dnuobs$ larger than 0.2\,$\mu$Hz,
this new treatment did not modify the first guess of the large
separations obtained in Section \ref{ogle-solar-like}. At very low
$\Dnuobs$, we noted a small change between the large separations
derived from the frequency difference between consecutive radial
orders or from the fit of Eq.~\refeq{tassoul_obs}. This change
increases when the large separation decreases, and reaches a value
of about 15\,\% at the lowest $\Dnuobs$.

This indicates that the almost constant large separation supposed
in Eq.~\refeq{tassoul_obs} is only an approximation. At very low
radial orders ($n\le3$), a gradient is observed that is larger
than the gradient inferred from the asymptotic expansion. This
gradient cannot be seen in \Kepler\ data because of the poorer
frequency resolution. The full characterization of the
parametrization of the spectrum at very low radial orders will
require a dedicated study, which is beyond the scope of this
paper.

\begin{figure*}
\centering
\includegraphics[width=14.8cm]{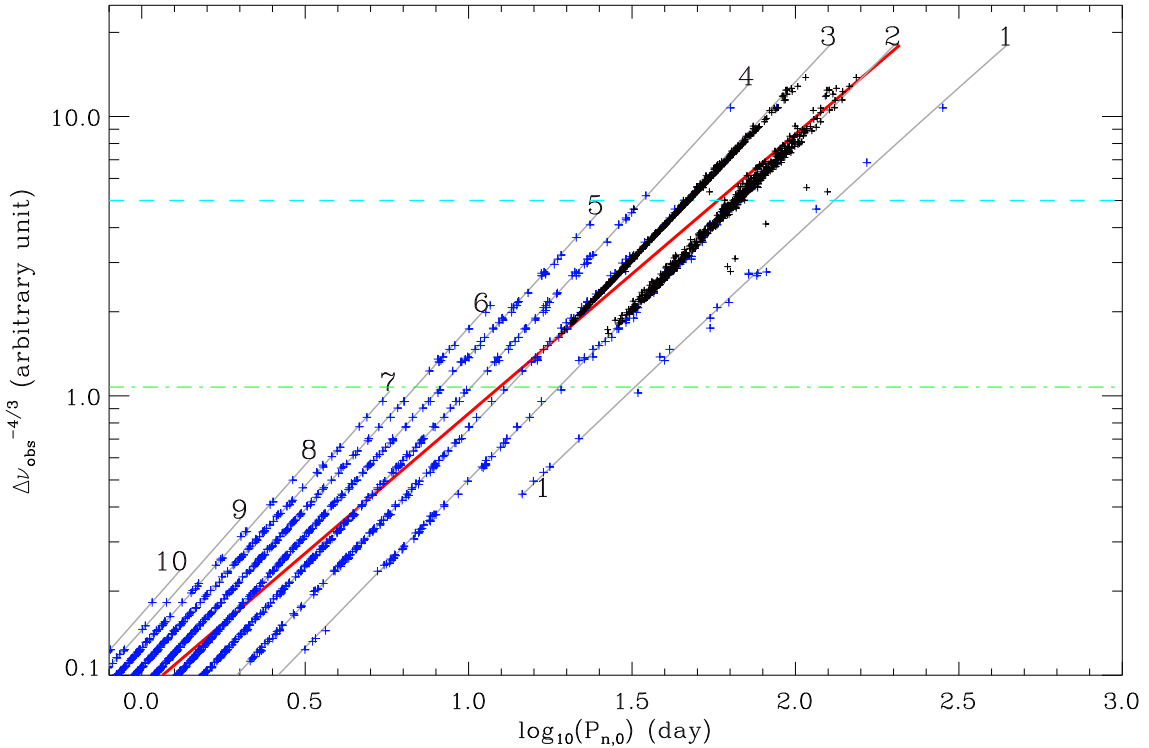}
\caption{PL relations of radial modes of AGB and RGB stars
observed with \Kepler\ (blue pluses) and OGLE (black symbols). The
proxy of the luminosity is derived from the estimate
$\Dnuobs^{-4/3}$. Each sequence, corresponding to a fixed radial
order, is fitted with the model provided by
Eq.~\refeq{tassoul_obs} (gray solid lines). The thick red line,
provided by $\numax \simeq 1/P$, indicates the the location of
ratio $\nmax = \numax/\Dnu - \epsobs$. The blue dashed line
corresponds to the tip of the RGB. The horizontal green dot-dashed
lines shows the region where the oscillation regime changes.
\label{PLdnu}}
\end{figure*}

\section{Identification of the period-luminosity sequences\label{PL-identification}}

Ground-based infrared survey results are most often presented with
PL sequences. In this section, we first aim to verify that the
solar-like oscillation pattern correspond to the PL sequences and
to provide an unambiguous identification of the observed
sequences. Second, we explore the consequence of this link.

\subsection{Seismic proxy of the luminosity}

Assessing PL relations requires determining the periods and the
luminosity. For field objects observed with \Kepler, determining
the luminosity can be done indirectly. A proxy of the luminosity
can be obtained from the black body relation. This requires the
use of effective temperature, provided by the \Kepler\ Input
Catalog \citep{2011AJ....142..112B}. From this catalog, we find
that $\Teff$ varies as $\Dnuobs^{0.068\pm0.011}$ for our cohort of
red giants.

From the black body relation and the $\numax$ scaling relation
\citep{2011A&A...530A.142B}, the luminosity scales as $L \propto
M\; \Teff^{7/2} \; \numax^{-1}$. Since $\numax$ suffers from large
uncertainties, we prefer to use the proxy
\begin{equation}\label{proxy_L}
  L \propto  M^{2/3}\; \Teff^{4} \; \Dnurho^{-4/3} ,
\end{equation}
with the dynamical frequency $\Dnurho$ for scaling as
$\sqrt{M/R^3}$. We prefer not to use the asymptotic large
separation $\Dnuas$, since an asymptotic value of the large
separation is not useful at such low orders.

Equation \refeq{proxy_L} is based on the assumption that $\numax
\propto \Dnurho^{3/4}$. At this stage, we are not able to relate
$\Dnurho$ and $\Dnuobs$, but first assume that they are
comparable. We have plotted $\Dnuobs^{-4/3}$ as a function of the
periods of the observed radial modes (Fig.~\ref{PLdnu}). Doing so
is in practice equivalent to supposing that all stars have the
same mass. This is justified because the oscillation pattern
hardly depends on the stellar mass \citep{2007MNRAS.378.1270X}.
For our set of data, we may assume that the mean stellar mass is
of about 1.3\,$M_\odot$, according to the mean value derived for
the cohort of red giants observed at higher $\numax$
\citep{2010A&A...522A...1K,2012A&A...540A.143M}.

\subsection{Identification of the sequences}

The identification of the PL relations (as in Fig. 1 of \SO) can
be obtained at this stage, following the stellar evolution from
the high-$\Dnuobs$ region, where pulsations are precisely
depicted, down to lower values where OSARG oscillations are seen.
The construction of a PL diagram with $\Dnuobs^{-4/3}$ as a proxy
of the luminosity provides a unique solution for identifying the
sequences observed in PL relations (Fig.~\ref{PLdnu}).

We note that, apart from a few points that represent fewer than
0.5\,\% of the measured periods, all periods fit the theoretical
ridges. The OGLE sequences $b_2$ and $b_3$ closely correspond to
the sequences with radial orders 2 and 3, respectively. This
proves that the identification proposed by DS10 is correct.
Oscillations in OSARG, and more generally in semi-regular
variables (SRV), are solar-like oscillations. A summary of the
identification of the sequences observed in SRV is given in Table
\ref{tab_equivalence}.

\begin{table}
 \caption{Equivalence between solar-like oscillations (labeled with the radial order $n$) and PL sequences}\label{tab_equivalence}
 \begin{tabular}{lll}
   \hline
$n$ & sequences & pulsations\\
   \hline
1 & $b_1$    & fundamental of  semi-regular variables (SRV)\\
2 & $b_2$    & 1st overtone of SRV\\
3 & $b_3$    & 2nd overtone of SRV\\
4 & ...      & ...               \\
\multicolumn{2}{l}{$\nmax \simeq  8$}& clump stars\\
\multicolumn{2}{l}{$\nmax \simeq 15$}& lower RGB\\
\multicolumn{2}{l}{$\nmax \simeq 22$}& Sun\\
   \hline
\end{tabular}
\end{table}

\subsection{Period-luminosity relation in the K band}

Usually, PL relations are expressed either with the brightness in
the K bandwidth  or with a Wesenheit index for avoiding reddening
issues \citep{1982ApJ...253..575M}. Here, we intend to calibrate
PL relations with the K magnitude. It depends on the effective
temperature, with an exponent $k =\dlog L\ind{K} / \dlog T$,
derived from the black body radiance, in the range [1.9, 2.1] for
the effective temperatures in the sample of stars (Table
\ref{bandeK}). We consider $\xKT \simeq 2.04$ as a reliable mean
value in the relation between the brightness $L\ind{K}$ in the K
band and the large separation $\Dnurho$. At fixed mass, we then
have from Eq.~\refeq{proxy_L}
\begin{equation}\label{lum-temp}
    \dlog L\ind{K} = -{4 \over 3}\  \dlog \Dnurho + \xKT \
    \dlog\Teff.
\end{equation}
Here, we need to introduce a relation between $\Dnuobs$ and
$\Dnurho$, which we empirically choose to write as a power law:
\begin{equation}\label{dnuas-dnuobs}
    \dlog \Dnurho = \coefDnu\ \dlog \Dnuobs .
\end{equation}
The exponent $\coefDnu$, presumably close to 1, remains unknown at
this stage. From the KIC temperatures \citep{2011AJ....142..112B},
we derive the mean relation between $\Teff$ and $\Dnuobs$:
\begin{equation}\label{teff-nuobs}
    \dlog\Teff = \coefT\ \dlog \Dnuobs ,
\end{equation}
with $\coefT$ about 0.068. Finally, we can derive the evolution of
the magnitude in the K band with the observed large separation:
\begin{equation}\label{lum-evolution}
    \diff K = 2.5 \left({4 \over 3}\  \coefDnu -  \xKT \coefT \right)
    \dlog\Dnuobs .
\end{equation}
This relation is established following the variation in the
observed large frequency separation $\Dnuobs$. It is therefore
representative of stellar evolution and does not correspond to any
PL sequence. In a next step, we have to retrieve the PL relation
for each observed sequence.

\begin{table}
  \centering
  \caption{Exponent $\xKT$ in the K band }\label{bandeK}
  \begin{tabular}{rcccc}
    \hline
$\Teff$& \multicolumn{3}{c}{$k (\lambda ) $}&$\xKT$\\
 (K)   &   2.0 &   2.2 &   2.4 &    \\
\hline
3800   & 2.23 & 2.10  & 1.99 & 2.10 \\
4000   & 2.16 & 2.03  & 1.93 & 2.04 \\
4200   & 2.09 & 1.97  & 1.88 & 1.98 \\
4700 (clump)  & 1.95 & 1.85  & 1.77 & 1.86 \\
 \hline
\end{tabular}
\end{table}

\subsection{Period-luminosity sequences in the K band}

The PL relations are a given set of different observed sequences,
each supposedly corresponding to a fixed radial order $n$. We note
$\Delta\Kn$ the variation in the magnitude along the sequence
corresponding to radial order $n$. We have to consider the
difference between the slopes of the PL sequences (at fixed radial
order) and the slope following stellar evolution (with the radial
orders $n$ of the observable modes evolving as $\nmax$). With the
proxy of the luminosity plotted in Fig.~\ref{PLdnu} and with
${\numax}^{-1}$ being considered as representative of the mean
period observed, we derive
\begin{equation}
 \nonumber \dlog P \simeq - \dlog \numax
   = - \coefDnuP\ {\dlog {\numax} \over\dlog \Dnuobs} \ \dlog P_n , \label{dif-pentes}
\end{equation}
where the derivatives $\coefDnuP$ of the relations $P_n =
1/\nu_{n,0}$, defined by
\begin{equation}\label{dnuobsPn}
    \dlog \Dnuobs = \coefDnuP\ \dlog P_n ,
\end{equation}
are estimated from Eq.~\refeq{tassoul_obs}. The relation between
stellar evolution and individual sequences is thus expressed by
\begin{equation}\label{dif-sequences-evolution}
    {\diff \Kn \over\dlog P_n} = - \coefDnuP\
    {\dlog\numax\over\dlog\Dnuobs}
     \left.{\diff K \over\dlog P}\right|_{\nmax} .
\end{equation}
From Eqs \refeq{lum-evolution}-\refeq{dif-sequences-evolution}, we
finally obtain the variation in the K magnitude with period on a
given sequence:
\begin{equation}\label{lum-dnu}
    \Delta\Kn \propto - 2.5\ \coefDnuP^2  \left({4\over 3} \ \coefDnu - \xKT \coefT \right)
    {\dlog\numax\over\dlog\Dnuobs} \
    \Delta\!\log P_n .
\end{equation}
With this result, it is now possible to interpret the slopes
$\Delta\Kn / \Delta\!\log P_n$ reported by \SO. In Table
\ref{tab_ridges}, we derive a mean value $\coefDnu$ for the
sequences $n=1$ to 3 observed by \SO\ in the Magellanic Clouds.

\begin{table}
 \caption{Sequences in the period-luminosity diagram}\label{tab_ridges}
 \begin{tabular}{rccc}
   \hline
radial order $n$   &   1    &   2    &   3       \\
slope $\coefDnuP$  &$-0.847$&$-0.902$&$-0.932$\\
   \hline
LMC sequence (\SO )& b$_1$   &  b$_2$ & b$_3$ \\
$\diff\Ks/\dlog P_n$& $-3.34$ & $-3.58$&$-3.72$\\
  $\gamma_n$          & 1.16    & 1.10   & 1.07 \\
    \hline
SMC sequence (\SO )& b$_1$   &  b$_2$ & b$_3$ \\
$\diff\Ks/\dlog P_n$& $-3.56$ & $-3.88$&$-4.40$\\
  $\gamma_n$          & 1.22    & 1.18   & 1.25 \\
\hline
\end{tabular}
\end{table}

\begin{table}
%
%
  \caption{Absolute seismic calibration}\label{rayon}
  \begin{tabular}{rrrrrrrrr}
    \hline
   $K$ &  $\numax$ & $\Dnuobs$ & $\Dnurho$& $\Teff$& $\gamma$ & $k$ &  $\Rrho$\\
       & ($\mu$Hz) & ($\mu$Hz) &($\mu$Hz)& (K)    &          &      & ($R_\odot$)\\
\hline
$-$8.55 &  0.05 & 0.024 & 0.016 &  3355 &  1.21 &  2.28 &  929\\
$-$8.06 &  0.07 & 0.033 & 0.023 &  3422 &  1.21 &  2.25 &  666\\
$-$7.58 &  0.11 & 0.044 & 0.033 &  3489 &  1.21 &  2.22 &  477\\
$-$7.10 &  0.15 & 0.059 & 0.047 &  3557 &  1.20 &  2.20 &  342\\
\multicolumn{8}{c}{\dotfill\ Tip of the RGB\ \dotfill}\\
$-$6.63 &  0.22 & 0.080 & 0.067 &  3626 &  1.19 &  2.17 &  246\\
$-$6.16 &  0.32 & 0.108 & 0.095 &  3695 &  1.17 &  2.14 &  177\\
$-$5.70 &  0.46 & 0.145 & 0.134 &  3763 &  1.14 &  2.12 &  128\\
$-$5.26 &  0.66 & 0.195 & 0.187 &  3833 &  1.11 &  2.09 & 94.9\\
$-$4.83 &  0.94 & 0.262 & 0.260 &  3905 &  1.08 &  2.07 & 71.4\\
$-$4.42 &  1.36 & 0.353 & 0.356 &  3979 &  1.05 &  2.05 & 55.2\\
$-$4.02 &  1.98 & 0.475 & 0.486 &  4057 &  1.03 &  2.02 & 43.7\\
$-$3.63 &  2.91 & 0.640 & 0.659 &  4138 &  1.02 &  2.00 & 35.2\\
$-$3.24 &  4.30 & 0.861 & 0.890 &  4223 &  1.01 &  1.97 & 28.8\\
$-$2.85 &  6.38 &  1.16 &  1.20 &  4309 &  1.01 &  1.95 & 23.7\\
$-$2.47 &  9.48 &  1.56 &  1.62 &  4398 &  1.00 &  1.93 & 19.6\\
$-$2.09 &  14.1 &  2.10 &  2.18 &  4489 &  1.00 &  1.91 & 16.2\\
$-$1.70 &  21.0 &  2.83 &  2.94 &  4583 &  1.00 &  1.88 & 13.5\\
$-$1.32 &  31.3 &  3.80 &  3.95 &  4678 &  1.00 &  1.86 & 11.2\\
    \hline
\end{tabular}

Variation with the K magnitude of the mean global seismic
parameters and of the parameters used for the relation between the
large separation $\Dnuobs$ and $\Dnurho$. $\Rrho$ is the stellar
radius derived from the scaling relation based on $\Dnurho$.

\end{table}

\begin{figure}
\includegraphics[width=8.8cm]{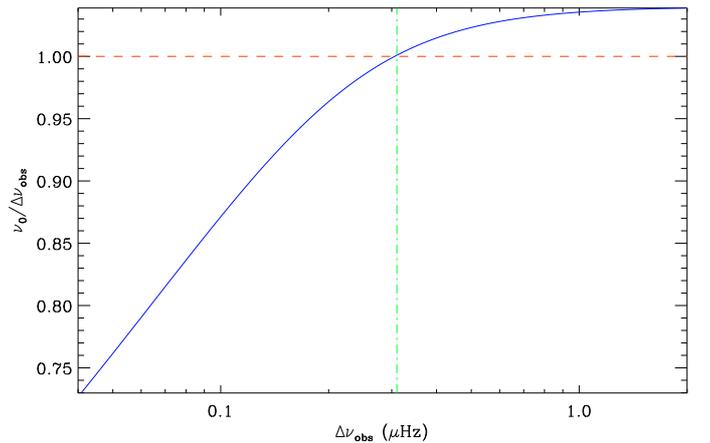}
\caption{Ratio $\Dnurho/ \Dnuobs$ as a function of $\Dnuobs$,
scaled according to the calibration derived from ground-based PL
relations. The dashed line indicates the 1:1 relation. The
vertical green dot-dashed lines shows the region where the
oscillation regime changes. \label{corres}}
\end{figure}

\section{Calibration of the period-luminosity relation\label{PL-calibration}}

From the identification of the sequences, we can derive the factor
$\gamma = \dlog\Dnurho / \dlog \Dnuobs$. Firstly, this opens the
possibility of measuring the mean stellar density from seismic
analysis in non-asymptotic conditions, under the hypothesis that
the scaling relation between $\numax$ and the acoustic cutoff
frequency is valid. Secondly, we can integrate
Eq.~\refeq{lum-evolution}.

\subsection{Calibration of the mean stellar density}

We have derived estimates $\gamma_n$ of $\gamma$ for each
sequence, according to Eq.~\refeq{lum-dnu}. Results are shown in
Table~\ref{tab_ridges}. We note a relative agreement better than
3\,\% between the coefficients $\gamma_n$ for the LMC and SMC
separately, even if we do not interpret the 7\,\% differences of
the slopes reported in the SMC and LMC by \SO.

Since the dependence of $\gamma_n$ in $n$ is less than the
uncertainties, we consider that the mean value of the $\gamma_n$
is representative of the scaling relation relating $\Dnurho$ to
$\Dnuobs$ at very low frequency:
\begin{equation}\label{coefdnu}
    \coefDnu \simeq 1.17\pm 0.08 .
\end{equation}
That the observed coefficient $\coefDnu$ is close to unity
indicates that $\Dnuobs$ provides a good proxy of a $\Dnurho$.
However, non-negligible divergence between these large separations
may appear when integrating Eq.~\refeq{dnuas-dnuobs} with the
result of Eq.~\refeq{coefdnu}.

We have built an empirical relation between the observed spacing
$\Dnuobs$ and the dynamical frequency $\Dnurho$, which assumes a
smoothly varying $\coefDnu$ coefficient between the different
regimes depending on stellar evolution. For the most-evolved
stages, at very low $\Dnuobs$, the value of $\coefDnu$ is given by
Eq.~\refeq{coefdnu}; for less-evolved stages, at larger $\Dnuobs$,
it corresponds to $\Dnurho \simeq \Dnuas \simeq 1.038\,\Dnuobs$
\citep{2013A&A...550A.126M}. The resulting model is shown in
Fig.~\ref{corres}. We have imposed the change of regime (i.e. a
smoothed change of $\coefDnu$) at $\Dnuobs=0.3\,\mu$Hz, in
agreement with the observation. This toy model is only a possible
solution, so it awaits a theoretical justification. We use it in
Section \ref{scalingDnu} to investigate the consequences of the
scaling provided by Eq.~\refeq{coefdnu}.

\begin{figure}
\includegraphics[width=8.8cm]{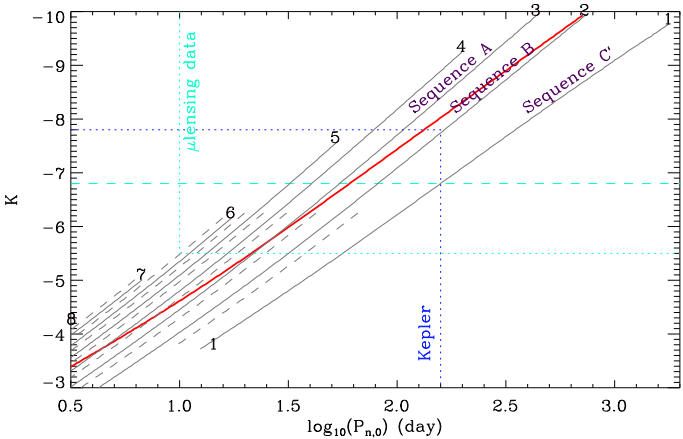}
\caption{Synthetic PL relations as proposed in Fig.~\ref{PLdnu},
but calibrated in K magnitude. Dipole modes (dashed lines) are
included for observed large separations down to $0.1\,\mu$Hz. PL
relations have been extrapolated towards bright magnitudes and
long periods. The dotted lines delimit the region of \Kepler\
observations with $\numax \ge 0.15\,\mu$Hz, and of OGLE
observations with $\numax \le 0.6\,\mu$Hz. Sequences A, B, and C'
identified at high luminosity by \SO\ are
indicated.\label{etalonnage_K_dnu}}
\end{figure}

\subsection{New calibration\label{absolutecalibration}}

With the determination of $\coefDnu$, it becomes possible to
integrate Eq.~\refeq{lum-evolution}, in order to relate the K
magnitude to the observed large separation. We have based the
calibration in $K$ on the sequences measured in the Large
Magellanic Cloud by \cite{2007AcA....57..201S}, taking the
distance modulus $\mu\ind{LMC} = 18.49$ into account
\citep{2013Natur.495...76P}. The variations in the parameter
$\gamma$ allow us to integrate the evolution  from the range where
OGLE measurements are made, encompassing the tip of the RGB
\citep[K magnitude of $-6.8$ in the solar neighborhood, according
to][]{2009ApJ...703L..72T}, to the red clump \citep[$K$ magnitude
about $-1.57$, according to][]{2007A&A...463..559V}.

Table \ref{rayon} shows different variables used for the
calibration: frequency $\numax$, effective temperature, and
parameters used for integrating Eq.~\refeq{lum-evolution}. We
estimate that the uncertainty of the calibration in $K$ is about
0.25. The PL sequences plotted in Fig.~\ref{etalonnage_K_dnu} take
this calibration into account. At low luminosity, sequences show a
smooth curvature, which is not depicted by the linear slope (in
log scale) reported by previous work since it occurs at periods
that are too short. In the OSARG domain, the sequences are drawn
by the decrease in the radial frequencies with stellar evolution.
When an RGB or AGB star evolves, its large separation decreases
and the observable radial orders decrease, too, from a mean value
$\nmax\simeq 8$ at the clump down to $\nmax\simeq 3$ at the tip of
the RGB, and down to $n=2$ or 1 for the most evolved semi-regular
variables. We note that the models of T13 do not reproduce this
scheme, since their theoretical sequences are not parallel to the
observed sequences.

At this stage, this seismic calibration is the same for RGB and
AGB stars, unlike classical PL relations
\citep{2003MNRAS.343L..79K}. Further observations of the red
giants at known distance will help improve the calibration, and
more modeling will help for understanding it.

\subsection{Scaling relations\label{scalingDnu}}

Previous asteroseismic work has shown that estimates of the
stellar mass and radius can be derived from seismic scaling
relations
\citep[e.g.,][]{2010A&A...522A...1K,2010A&A...517A..22M}.
Continuing efforts are being made for improving the accuracy of
these relations: calibration with grid-based modeling
\citep{2011ApJ...743..161W}, correction for evolution
\citep{2012MNRAS.419.2077M}, correction of bias, and calibration
resulting from the comparison of modeled stars and second-order
asymptotic information \citep{2013A&A...550A.126M}.

In this work, we propose a similar scaling relation, but derived
from the determination of $\Dnurho$ provided by \Kepler\ and
ground-based observations. Results for the variation in the mean
stellar radius with evolution are shown in Table~\ref{rayon}. The
ratio $\Dnurho / \Dnuobs$, which is supposed to be close to 1.04
in the red giant regime corresponding to the lower RGB
\citep{2013A&A...550A.126M}, decreases significantly when the
frequency decreases. As a consequence, scaling relations using
$\Dnuobs$ are only approximate. Further modeling, similar to
\cite{2011ApJ...743..161W}, will help for investigating this in
more detail and for tightening the link between $\Dnuobs$  and
$\Dnurho$.

\section{Discussion\label{discussion}}

In this section, we intend to show how the stochastic nature of
solar-like oscillations helps for understanding some
characteristics of PL features. Firstly, the amplitudes of the
oscillations can explain the emergence of significant mass loss.
More speculative results concern the determination of the stellar
evolutionary status and the nature of the change of regime
detected below the tip of the RGB. Last but not least, we
investigate how this work and all the information compiled by PL
analysis can interact to provide refined distance scales.

\subsection{Stochastic oscillations versus Mira variability}

With statistical arguments, \cite{2001ApJ...562L.141C} have
suggested that oscillations in semi-regular variables are
stochastically excited. A similar conclusion was reached by
\cite{2005MNRAS.361.1375B} for the star $L_2$ Puppis. Further work
by DS10 and T13 has shown that PL relations are drawn by
stochastically excited oscillations. Here, the identification of
PL relations with solar-like oscillations confirms this. The
measurements of complex visibility function, with clear
contributions of non-radial modes, reinforce this interpretation.

Semi-regular variables are red giants on the RGB or AGB.
Semi-regularity is due to the small number of stochastically
excited oscillation modes that are observed. Their complex beating
and evolution with time are enough for explaining the changes. The
scaling relation of $\Amax$ (Eq.~\ref{eqt-Amax}), translated in
magnitude variation, fits with microlensing results
\citep[e.g.,][]{2006ApJ...650L..55D,2010MNRAS.405.1770N}. We note
that variations of about one magnitude for oscillation periods
about 500 days are compatible with the extrapolation of the
scaling relation of $\Amax$. However, much larger amplitudes, as
reported by \cite{2005AJ....129..768F}, cannot be explained.

As expected, the Mira variability, with amplitudes of a few
magnitudes, cannot be interpreted in terms of solar-like
oscillation pressure modes. As stars in the instability strip, the
Mira stars seem to obey a specific oscillation pattern, which does
not correspond to solar-like oscillations. At this stage, it is
not possible to provide an unambiguous identification of the
radial order and degree of sequences C and D identified by
ground-based measurements (\SO ).

\subsection{Perturbation to hydrostatic equilibrium and mass loss\label{mass_loss}}

The measurement of the maximum amplitude $\Amax$ has shown that
high values are attained when stars climb the AGB. The link
between the maximum amplitude and the relative displacement $\dR
/R$ helps us understand the result of \cite{2009MNRAS.395L..11G},
who state that massive mass loss is seen for main periods longer
than 60 days (corresponding to $\numax \simeq  0.2\,\mu$Hz).

The surface gravity can be expressed as a function of the large separation as
\begin{equation}\label{gravitation}
    g = {\mathcal{G} M \over R^2} \simeq (\alpha \Dnuobs)^2 \ R ,
\end{equation}
with $\alpha \simeq 3.6$ according to a mean estimate derived from
our work. The acceleration of the oscillation associated to a
displacement $\dR$ is related to $\numax$ by
\begin{equation}\label{acceleration}
    a = (2\pi \numax)^2 \ \dR .
\end{equation}
Equality between $a$ and $g$ is reached when
\begin{equation}\label{egalite}
   {\dR \over R} \simeq {0.36 \over \nmax^2} .
\end{equation}

In the K band, the relative brightness fluctuations depend on the
relative Lagrangian radius and temperature variations
\begin{equation}\label{lum-temp-rayon}
    {\delta L\ind{K}\over L\ind{K}}=2\ {\dR \over R} + \xKT \ {\delta \Teff \over
    \Teff}.
\end{equation}
As seen in Section~\ref{ampmax}, the maximum value of $\delta
L\ind{K} / L\ind{K}$ corresponds to the maximum amplitude $\Amax$
derived from the power spectrum. In adiabatic conditions, the
relative brightness variation reduces to the temperature
contribution. This results from the fact that the relative radius
and effective temperature Lagrangian changes are related by
\begin{equation}\label{dlum}
   {\dR \over H\ind{p}} = - {\Gamma_1 \over \Gamma_1 -1} \ {\delta \Teff \over
   \Teff},
\end{equation}
where $\Gamma_1$ is the local adiabatic exponent, and $H\ind{p}$
the pressure scale height \citep[e.g.,][]{1995A&A...293..586M}. In
a spherically symmetric unperturbed star, the pressure scale
height in the photosphere  is much less than the stellar radius,
which explains that  $\Amax \simeq \xKT \,  | \delta \Teff /
\Teff|$. However, in conditions where $\Teff$ presents important
relative variation, hydrostatic equilibrium may not be met in the
upper envelope, so that Eq.~\refeq{dlum} may be not valid.

We therefore investigated the case where the term ${\dR / R}$
significantly contributes to $\Amax$. The relation $\Amax = 2\, |
{\dR / R}|$, superimposed on the observed amplitudes
(Fig.~\ref{fig-Amax}), then helps to quantify the results of
\cite{2009MNRAS.395L..11G}: the observed maximum amplitude
corresponding to $a=g$ occurs for $\numax\simeq 0.2\,\mu$Hz.  The
quantitative coincidence is puzzling and shows that the possible
disruption of the external layers of evolved red giants has to be
investigated. Qualitatively, efficient mass loss certainly occurs
when the oscillation amplitude in the external layers is large
enough for imposing an oscillation acceleration comparable to the
surface gravity. As a result, the uppermost regions are no longer
firmly linked to the envelope and can be easily ejected.

\subsection{AGB versus RGB\label{AGBRGB}}

In \SO, sequences labeled with $a$ (respectively $b$) correspond
to stars on the AGB (respectively on the RGB). We tried to test
different ways able to similarly disentangle AGB from RGB stars
with \Kepler\ asteroseismic data.

\cite{2011Natur.471..608B} and \cite{2011A&A...532A..86M} have
shown that mixed modes provide a discriminating test between RGB
and clump stars. Unfortunately, the impossibility of measuring
mixed-mode spacings at low $\Dnuobs$ precludes a similar use for
identifying the evolutionary status. Independent of the mixed mode
pattern, the exact location of the radial modes also helps
distinguish RGB from red clump stars \citep{2012A&A...541A..51K}.
At the moment, we lack information on the fine structure of the
oscillation spectra in the upper RGB to perform a similar
investigation. Because AGB stars have higher $\Teff$ than RGB
stars, we also tried to compare the oscillation patterns depending
on $\Teff$, but failed to identify firm signatures.

At this stage, we are left with a degeneracy between the
oscillation spectra of AGB and RGB stars observed with \Kepler .
The comparison with ground-based observations will help remove it.

\subsection{Regime change}

We have seen many changes occurring around $\numax \simeq
1\,\mu$Hz. At this stage, we have no definitive explanation. We
may imagine that non-linear effects become strong enough when
$\Amax$ is larger than $10^{-3}$. Alternatively, if a majority of
AGB stars are seen at this stage, this could be the signature of
the third dredge-up. Such a signature could explain the lack of
stars, since the input of heavy elements in the envelope may have
boosted the evolution along the AGB. Again, the wealth of
information obtained with ground-based surveys will help
ascertain this.

\subsection{Distance scale and microlensing surveys}

The parametrization of the solar-like oscillation spectrum at low
frequency and the measurement of the mode visibility developed for
the \Kepler\ giants, corrected for the gradient seen at very low
$\Dnuobs$,  can be used to analyze ground-based data from a
microlensing survey. Identifying pulsations in evolved red giants
with solar-like oscillations will help refine our understanding
of the distance measurements provided by the PL relations
\citep[e.g.,][]{2004ASPC..310..304F}, since the analysis can now
stand on a firm basis.

Periods can be precisely measured and interpreted with more
accurate PL relations. Asteroseismology is in fact unable to
provide the zero point of the PL relations, but can provide
accurate slopes that vary with $\numax$
(Fig.~\ref{etalonnage_K_dnu}). With the explanation proposed by
our \Kepler\ measurements, the difference in PL slopes between the
LMC and the SMC is surprising (Table~\ref{tab_ridges}): we do not
expect the influence to the metallicity to be important.
Observationally, PL relations do not depend on metallicity, as
derived from the observation in a large number of infrared
wavelengths \citep{2009A&A...495..157S}. This result is fully
compatible with solar-like oscillations. Theoretically,
simulations of non-adiabatic oscillations in red giants have shown
that the oscillation hardly depends on mass and metallicity
\citep{2007MNRAS.378.1270X}. In fact, metallicity acts on
$\numax$, hence on the luminosity, essentially through the
influence of the Mach number
\citep[e.g.,][]{2010A&A...509A..15S,2011A&A...530A.142B}. However,
recent work shows that the $\numax$ -- $\nu\ind{c}$ relation
hardly depends on the Mach number for red giants \citep{belkacem},
which lowers the influence of metallicity.

Analyzing the microlensing data again, taking the analytical
description of the oscillation pattern depicted by
Eq.~\refeq{tassoul_obs} into account, will be fruitful, since the
refined seismic analysis will benefit from microlensing
information. As a consequence, population studies and distance
measurements can be boosted with the new tool and paradigm
proposed by our work.

\section{Conclusion\label{conclusion}}


We have confirmed that PL relations in evolved red giants are due
to solar-like oscillations. This was done with an analytical
description of the oscillation spectra. We calibrated this
description with red giants observed with \Kepler\ compared to
similar stars observed in ground-based microlensing surveys.

Interpreting variability in terms of pressure modes helps clarify
many issues:
\begin{itemize}
    \item confirmation of the excitation mechanism (solar-like
    oscillations) in the semi-regular variables;
    \item identification of the different sequences of the PL
    relations;
    \item identification of both radial and non-radial modes, except at
    very low $\numax$ where radial modes are dominant;
    \item scaling relations compatible with
    a magnitude variation $\Delta m =1$ for periods of 500 days;
    \item possible identification of the process explaining the massive mass-loss
    starting for periods longer than 60 days;
    \item justification of the independence of the PL relations on metallicity.
\end{itemize}
However, we failed at this stage to unambiguously disentangle RGB from AGB stars.

An important physical output of calibrating the PL relation
consists in the significant difference between the observed large
frequency separation and the dynamical frequency proportional to
the square root of the mean stellar density. We have proven that
these frequencies differ and have provided a toy model for
relating them. For the most evolved red giants, the stellar mean
density is higher than derived from a flat scaling with the
observed large separation. As a consequence, the stellar radius is
smaller than extrapolated from the usual scaling relation.

\begin{acknowledgements}

Funding for the Discovery mission \Kepler\ is provided by NASA's
Science Mission Directorate.  KB, MJG, EM, BM, RS and MV
acknowledge financial support from the ``Programme National de
Physique Stellaire" (PNPS, INSU, France) of CNRS/INSU and from the
ANR program IDEE ``Interaction Des \'Etoiles et des
Exoplan\`etes'' (Agence Nationale de la Recherche, France). SH
acknowledges financial support from the Netherlands Organisation
for Scientific Research (NWO). WAD is supported by Polish NCN
grant DEC-2012/05/B/ST9/03932. YE acknowledges support from STFC
(The Science and Technology Facilities Council, UK). This work
partially used data analyzed under the NASA grant NNX12AE17GPGB
and under the European Community's Seventh Framework Program grant
(FP7/2007-2013)/ERC grant agreement n. PROSPERITY.

\end{acknowledgements}

\bibliographystyle{aa} 
\bibliography{biblio_low}

\listofobjects
\end{document}